\documentclass[conference]{IEEEtran}

\IEEEoverridecommandlockouts
\usepackage{amsmath,amssymb,amsfonts}
\usepackage{algorithmic}
\usepackage{graphicx}
\usepackage{textcomp}
\usepackage{xcolor}
\usepackage{subcaption}
\usepackage{comment}

\begin{document}

\title{From Spaceborne to Airborne: SAR Image Synthesis Using Foundation Models for Multi-Scale Adaptation\\}

\author{

\IEEEauthorblockN{1\textsuperscript{st} Solène Debuysère}
\IEEEauthorblockA{\textit{ONERA - DEMR} \\
Université Paris Saclay \\
Palaiseau, France \\
solene.debuysere@onera.fr}
\and
\IEEEauthorblockN{2\textsuperscript{nd} Nicolas Trouvé}
\IEEEauthorblockA{\textit{ONERA - DEMR} \\
Université Paris Saclay \\
Palaiseau, France \\
nicolas.trouve@onera.fr}
\and
\IEEEauthorblockN{3\textsuperscript{th} Nathan Letheule}
\IEEEauthorblockA{\textit{ONERA - DEMR} \\
Université Paris Saclay \\
Palaiseau, France \\
nathan.letheule@onera.fr}
\and
\IEEEauthorblockN{4\textsuperscript{th} Olivier Lévêque}
\IEEEauthorblockA{\textit{ONERA - DEMR} \\
Université Paris Saclay \\
Palaiseau, France \\
olivier.leveque@onera.fr}
\and
\centering
\IEEEauthorblockN{5\textsuperscript{th} Elise Colin}
\IEEEauthorblockA{\textit{ONERA - DTIS} \\
Université Paris Saclay \\
Palaiseau, France \\
elise.colin@onera.fr}

}

\maketitle

\begin{abstract}

The availability of Synthetic Aperture Radar (SAR) satellite imagery has increased considerably in recent years, with datasets commercially available. However, the acquisition of high-resolution SAR images in airborne configurations, remains costly and limited. Thus, the lack of open source, well-labeled, or easily exploitable SAR text-image datasets is a barrier to the use of existing foundation models in remote sensing applications. In this context, synthetic image generation is a promising solution to augment this scarce data, enabling a broader range of applications. Leveraging over 15 years of ONERA's extensive archival airborn data from acquisition campaigns, we created a comprehensive training dataset of 110 thousands SAR images to exploit a 3.5 billion parameters pre-trained latent diffusion model \cite{Baqu2019SethiR}. In this work, we present a novel approach utilizing spatial conditioning techniques within a foundation model to transform satellite SAR imagery into airborne SAR representations. Additionally, we demonstrate that our pipeline is effective for bridging the realism of simulated images generated by ONERA's physics-based simulator EMPRISE \cite{empriseem_ai_images}. Our method explores a key application of AI in advancing SAR imaging technology. To the best of our knowledge, we are the first to introduce this approach in the literature.  


\end{abstract}

\section{Introduction}

Among existing generative models, diffusion-based models have emerged as a powerful solution to address key limitations of other auto-regressive, flow-based or implicit models approaches. Indeed, models such as DDPMs \cite{ho2020denoising} or GANs \cite{goodfellow2014generative} prove to be limiting in the diversity or quality of scenes that can be generated. On the other hand, by taking random noise and text prompts as inputs, conditionnal diffusion models such as Stable Diffusion \cite{rombach2022highresolution}, Flux, Imagen, and DALLE- 3 have made a significant impact on text to-image generation. And by projecting data into latent spaces, Stable Diffusion offers computational advantages that streamline training and sampling. In that context, the use of these pre-trained foundation models offers a great opportunity for generating images due to abilities in modeling image details with high-level semantics. However, Synthetic Aperture Radar (SAR) imagery, with its unique statistical properties—such as speckle noise, high dynamic range, and geometric distortions poses challenges that traditional vision models, primarily designed for optical images, struggle to address. This limitation complicates both the training process and the precise control of the model output, highlighting a significant avenue for research.  Recently, Trouvé and al. \cite{trouve2024synthesis} demonstrated the feasibility of generating simulated radar images by conditioning the Stable Diffusion model on textual prompts. They employed Stable Diffusion v1.5 (860 million parameters) 
\cite{rombach2022highresolution} on a dataset of 10,000 samples from ONERA's archives. Lately, we \cite{debuysere} presented a conditional multi-resolution latent diffusion pipeline that expand SAR images from a size of 512 × 512 to 2048 × 2048, progressively improving resolution through three stages: 160 cm, 80 cm, and finally 40 cm \cite{debuysere}.
Here, our study leverages Stable Diffusion XL (3.5 billion parameters) \cite{podell2023sdxl} to transform satellite SAR imagery into airborne SAR representations. 


Tackling this kind of model has been made possible by creating an extensive training dataset (110K), developed using archival data from ONERA's airbone sensor. We used a 15-year collection of X-band SAR images from southern France, obtained via a Falcon 20 aircraft with the SETHI system. We selected 461 large SAR images from different acquisition campaigns with varied acquisition parameters and filtering for quality and metadata. The data were processed and segmented, resulting in over 100K image samples of one megapixel at resolutions of 40 cm.

First, we introduce the architecture of Stable Diffusion and the spatial conditioning module: ControlNet \cite{zhang2023adding}. Finally, we present our diffusion latent upscaling pipeline with Stable Diffusion XL for satellite resolution refinement through ControlNet modules. We aim to creativly enhance TerraSAR-X image resolution. An additional application consists in enhancing the realism of simulated images generated by ONERA's physics-based simulator. 


\section{Methodology}
\subsection{Objectives}
First, our work included the creation of a diverse training dataset that implied spectral recentering, subsampling, calibration, standardization of the data dynamics and errors detection. In addition to SAR data preparation, the project involved pairing SAR and optical images and ensuring geographic overlap. Then, we captioned pairs with optical images (from open-source IGN datasets) with CogVLM2 model. We first worked on a dataset of 10.000 samples at 80cm resolution and 40cm resolution. Recently, a significant achievement was scaling data processing 20 times compared to earlier efforts. This work led to the creation of 100,000 SAR-optical pairs at a 40cm resolution. 
Our goal is to fine-tune the pre-trained foundation model Stable Diffusion on this dataset. Then, its application consists in developing a creative latent upscaling pipeline by incorporating two ControlNet modules operating at different resolutions within Stable Diffusion XL (SDXL).

\subsection{Architecture: Stable Diffusion Foundation Model}
\paragraph{Inputs} The model learns in a latent space rather than in pixel space. Indeed, the texts - images pairs inputs are compressed into a more compact representation separately. First, a Variational Autoencoder (VAE) compresses images into a more compact representation. The VAE encodes an image $x$ into a latent representation $z$:

\begin{equation}
    z = E(x), \quad \tilde{x} = D(z)
\end{equation}

where $E$ is the encoder, $D$ is the decoder, and $z$ is the latent representation of image $x$. Images are compressed by a factor of 8 to obtain $z$.
Secondly, the text prompts inputs $y$ - description of the scene we want to generate - are converted into embeddings $\tau_{\theta}(y)$ via CLIP's text encoder, resulting in a matrix where each token (word) is a high-dimensional vector. The model is limited to 75 words plus 2 tokens for start and end, leading to 77 token embeddings.

\paragraph{Training Process} The model learns by destroying training data, or compressed images, through the successive addition of Gaussian noise (forward process), and then training a model to reverse this process (reverse process) and recover the original data. The generation is performed by reversing this process, to turn noise back into data.

Specifically, the forward diffusion process which gradually corrupts an input data $x_0$ over $T$ timesteps by adding Gaussian noise is defined as:
\begin{equation}
z_t = \sqrt{\bar{\alpha}_t} \, z_0 + \sqrt{1 - \bar{\alpha}_t} \, \epsilon, \quad \epsilon \sim \mathcal{N}(0, \mathbf{I})
\end{equation}
where $\bar{\alpha}_t = \prod_{s=1}^t \alpha_s$ with $\alpha_t = 1 - \beta_t$ represent the cumulative product of noise reduction at each timestep.
Similar to a Markov chain, it is assumed that the distribution at each timestep depends solely on the previous timestep, leading to:
\begin{equation}
q(z_t | z_{t-1}) := \mathcal{N}(z_t; \sqrt{1 - \beta_t} \, z_{t-1}, \beta_t \, \mathbf{I})
\end{equation}
The reverse diffusion process denoises the latent data step by step. In particular, the backbone model predicts the mean $\mu_\theta$ and variance $\Sigma_\theta$ at each timestep to reverse the noise. This process is guided by the text embeddings derived from CLIP Text Encoder to obtain $\tau_{\theta}(y)$:
\begin{equation}
    p_{\theta}(\mathbf{z}_{t-1} | \mathbf{z}_t, y) := \mathcal{N}(\mathbf{z}_{t-1}; \mu_{\theta}(\mathbf{z}_t, t, \tau_{\theta}(y)), \Sigma_{\theta}(\mathbf{z}_t, t))
\end{equation}

The loss function used to train the model minimizes the difference between the true noise $\epsilon$ and the predicted noise $\epsilon_\theta$:
\begin{equation}
    L_{\text{cond}} = \mathbb{E}_{t, \mathbf{z}_0, \epsilon, y} \left[ \left\| \epsilon - \epsilon_{\theta}(\mathbf{z}_t, t, \tau_{\theta}(y)) \right\|^2 \right]
\end{equation}


\subsection{Strategy: Transfer learning in Radar domain}
Fine-tuning Stable Diffusion for SAR image generation represents a significant challenge, as these models have been trained on billions of optical images and lack prior exposure to SAR data. And SAR images have distinct textural patterns and speckle noise, which differ from natural images. This makes it difficult for the model to learn the correct texture patterns during generation. Thus, we used a Low-Rank Adaptation (LoRA) fine-tuning pipeline of Stable Diffusion XL to only train a subset of parameters from the backbone model, to reduce the risk of overfitting or collapse of the model. We fine-tuned the model on three different resolutions: 160cm, 80cm and 40cm. 

\subsection{Spatial conditioning: ControlNet Module}

To incorporate spatial conditioning through ControlNet, we use a spatial condition term $\phi_{\theta}(x)$ extracted from the input image $x$. This spatial information (e.g., edges or textures) complements the text condition during denoising. The reverse diffusion process becomes:

\begin{equation}
    p_{\theta}(\mathbf{z}_{t-1} | \mathbf{z}_t, y, x) = \mathcal{N}\left(\mathbf{z}_{t-1}; \mu_{\theta}(\mathbf{z}_t, t, \tau_{\theta}(y), \phi_{\theta}(x)), \Sigma_{\theta}(\mathbf{z}_t, t)\right)
\end{equation}

Here, the mean prediction $\mu_{\theta}$ is conditioned on the latent variable $\mathbf{z}_t$, the timestep $t$, the text embedding $\tau_{\theta}(y)$, and the spatial condition $\phi_{\theta}(x)$ derived from ControlNet.

\section{Applications}
\subsection{Objectives}
Our goal is to enhance the realism of simulated images generated by ONERA's physics-based simulator and improve TerraSAR-X image resolution with a multi-resolution pipeline through ControlNet modules in Stable Diffusion XL (SDXL).
\subsection{Pipeline architecture}
\begin{figure}
    \includegraphics[width=0.45\textwidth]{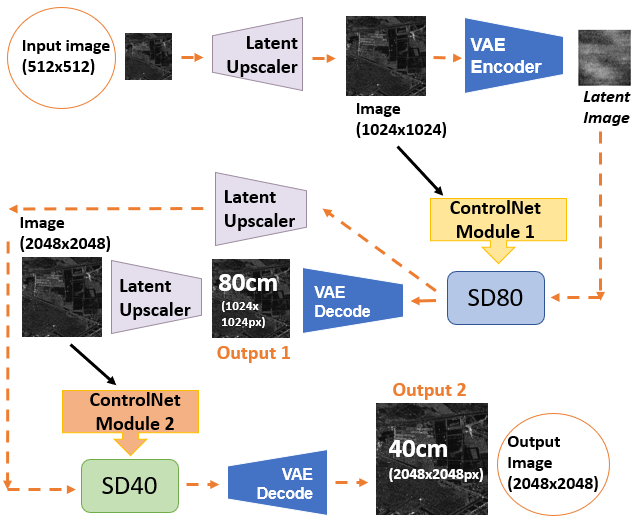}
    \caption{Creative upscaling pipeline}
     \label{fig:pipeline}
\end{figure}

\begin{figure*}
    \centering
    \begin{subfigure}{0.19\textwidth}
        \includegraphics[width=\textwidth]{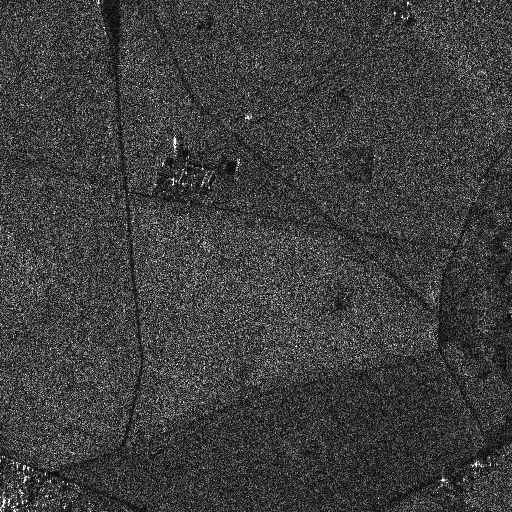}
        \caption{}
    \end{subfigure}\hfill
    \begin{subfigure}{0.19\textwidth}
        \includegraphics[width=\textwidth]{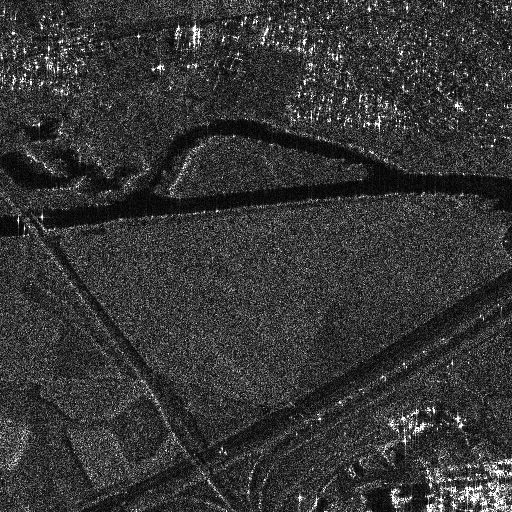}
        \caption{}
    \end{subfigure}\hfill
    \begin{subfigure}{0.19\textwidth}
        \includegraphics[width=\textwidth]{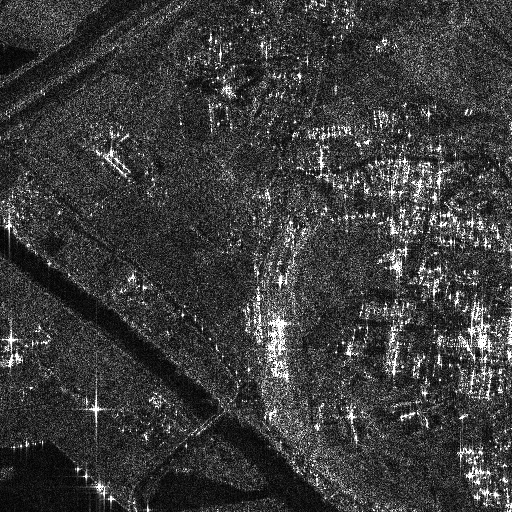}
        \caption{}
    \end{subfigure}\hfill
    \begin{subfigure}{0.19\textwidth}
        \includegraphics[width=\textwidth]{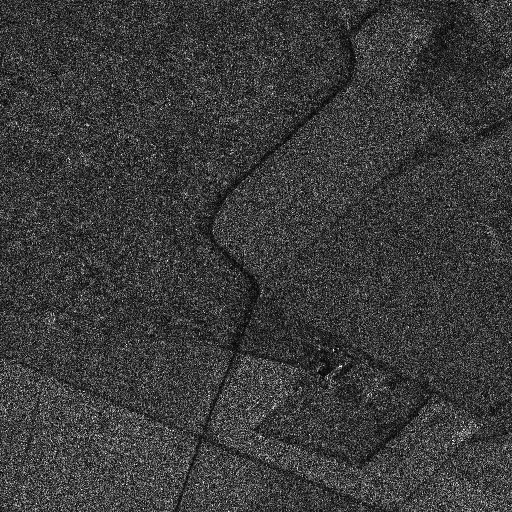}
        \caption{}
    \end{subfigure}\hfill
    \begin{subfigure}{0.19\textwidth}
        \includegraphics[width=\textwidth]{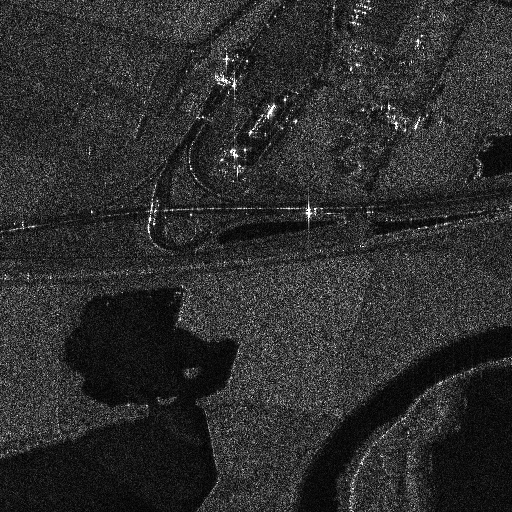}
        \caption{}
    
    \end{subfigure}

    \vspace{5pt}
    \begin{subfigure}{0.19\textwidth}
        \includegraphics[width=\textwidth]{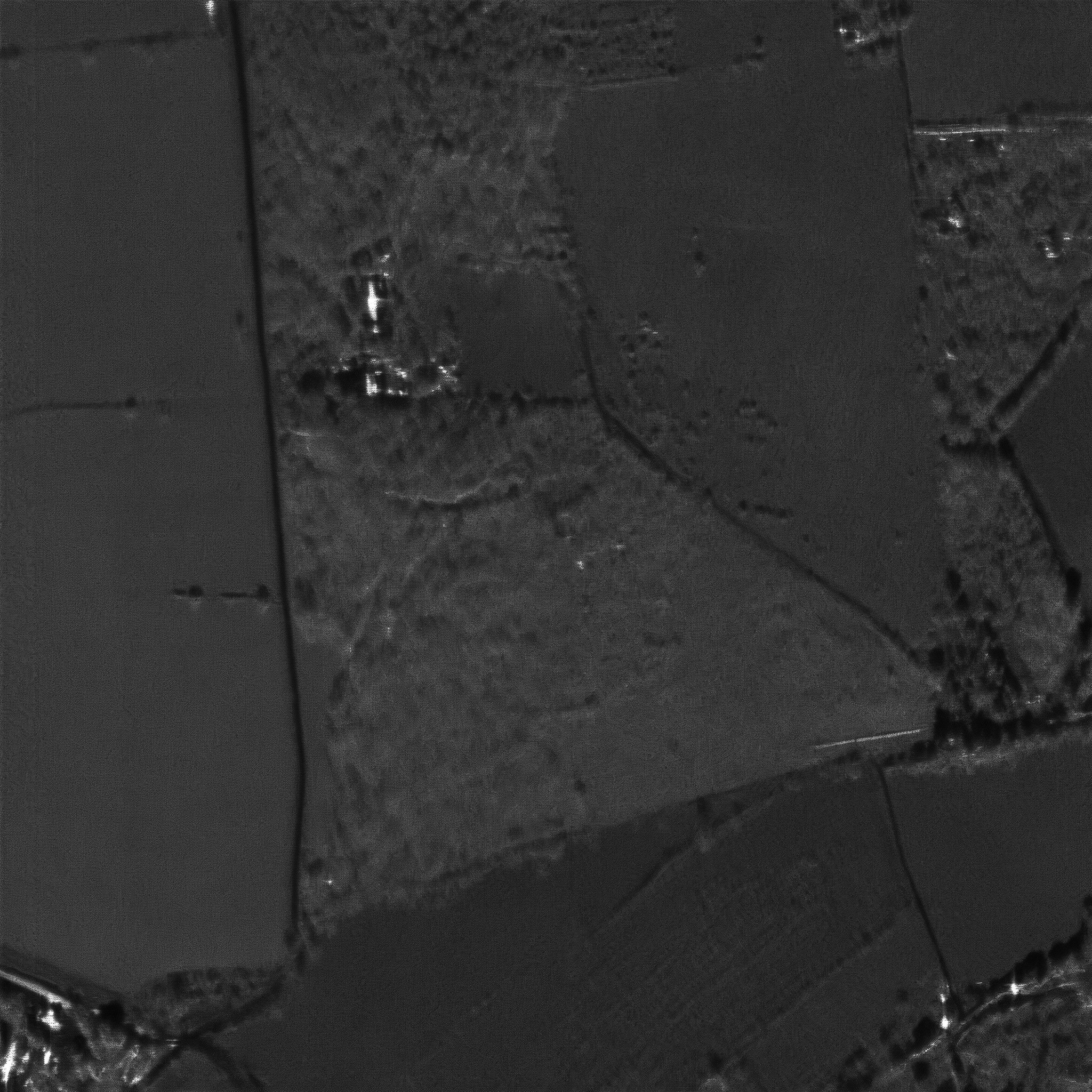}
        \caption{}
    \end{subfigure}\hfill
    \begin{subfigure}{0.19\textwidth}
        \includegraphics[width=\textwidth]{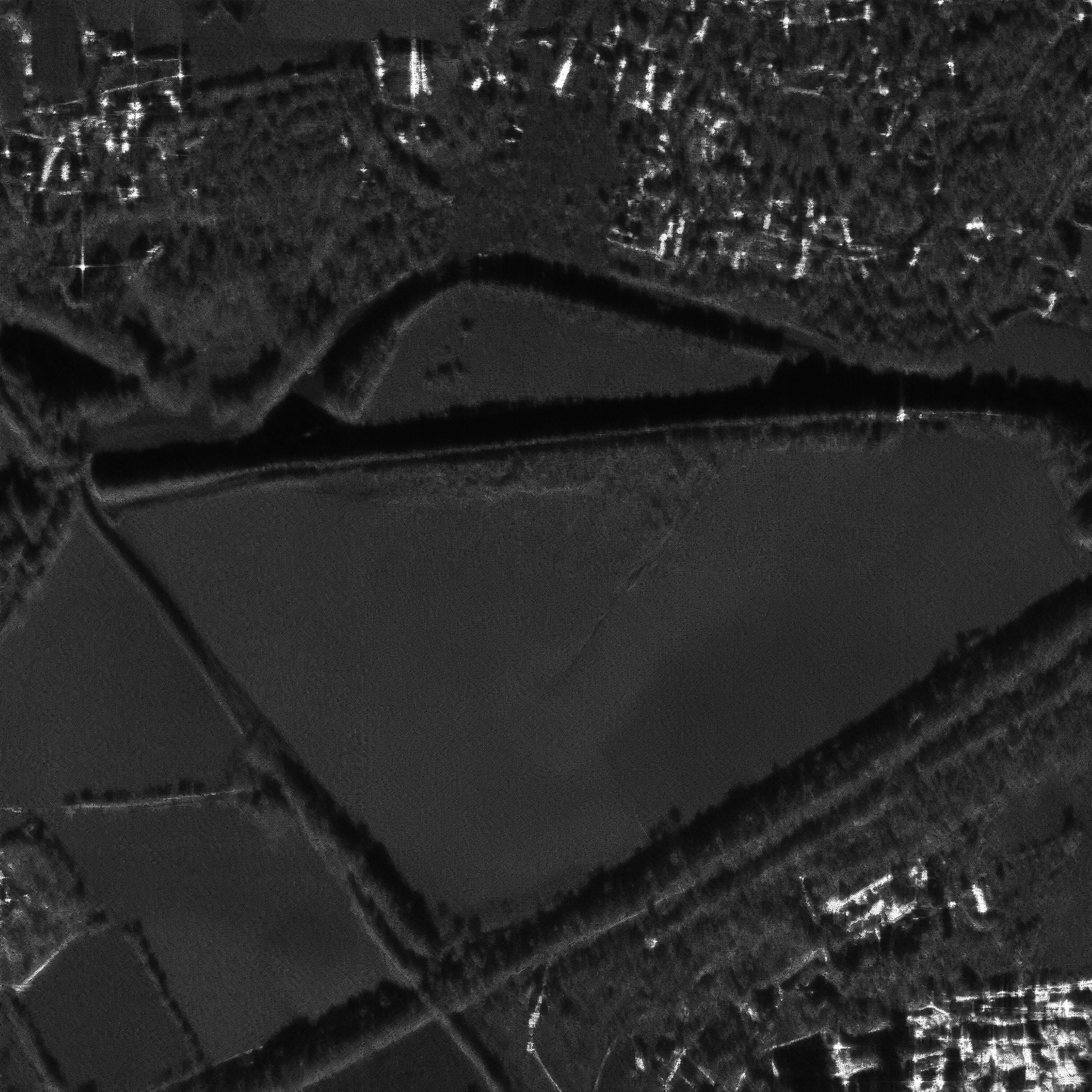}
        \caption{}
    \end{subfigure}\hfill
    \begin{subfigure}{0.19\textwidth}
        \includegraphics[width=\textwidth]{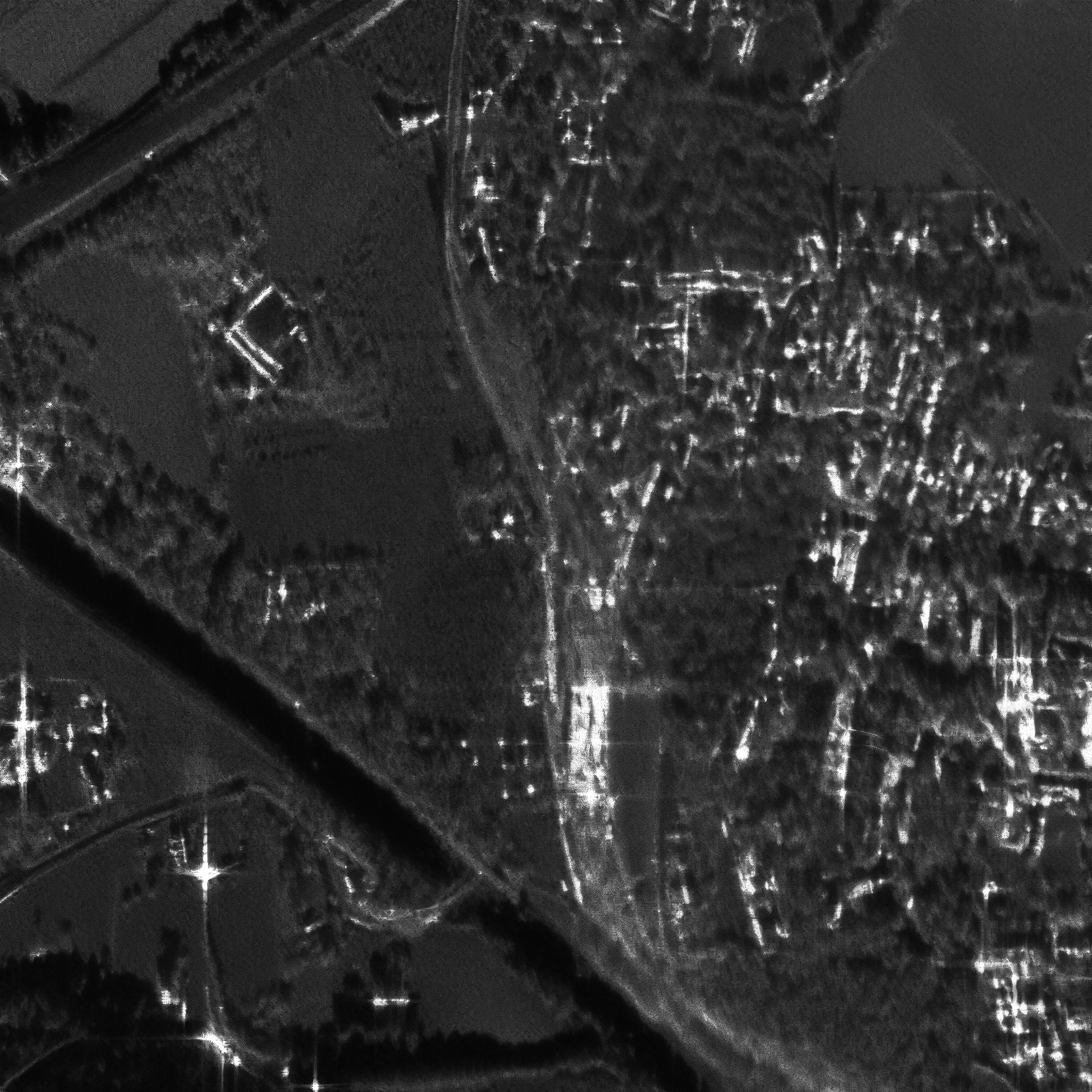}
        \caption{}
    \end{subfigure}\hfill
    \begin{subfigure}{0.19\textwidth}
        \includegraphics[width=\textwidth]{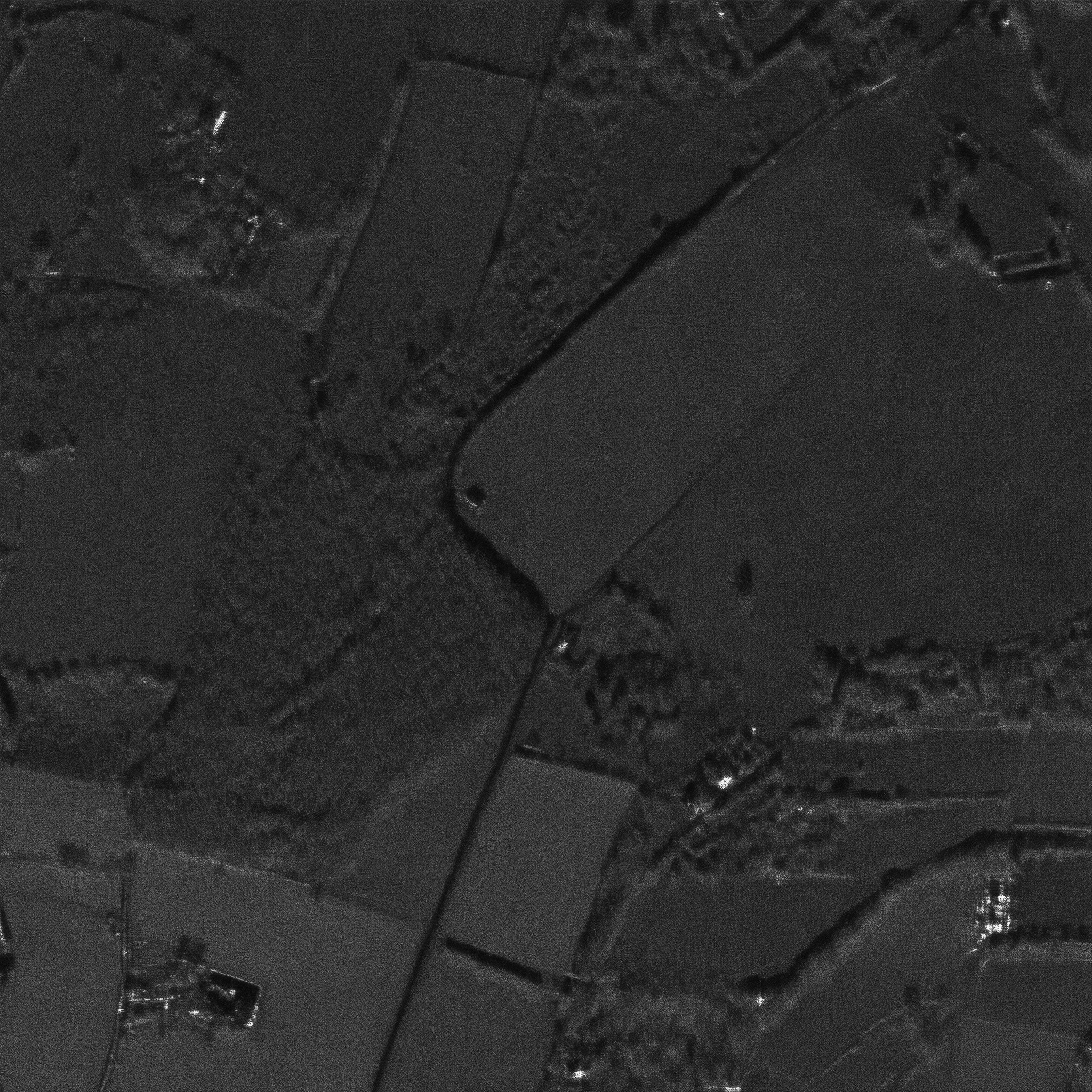}
        \caption{}
    \end{subfigure}\hfill
    \begin{subfigure}{0.19\textwidth}
        \includegraphics[width=\textwidth]{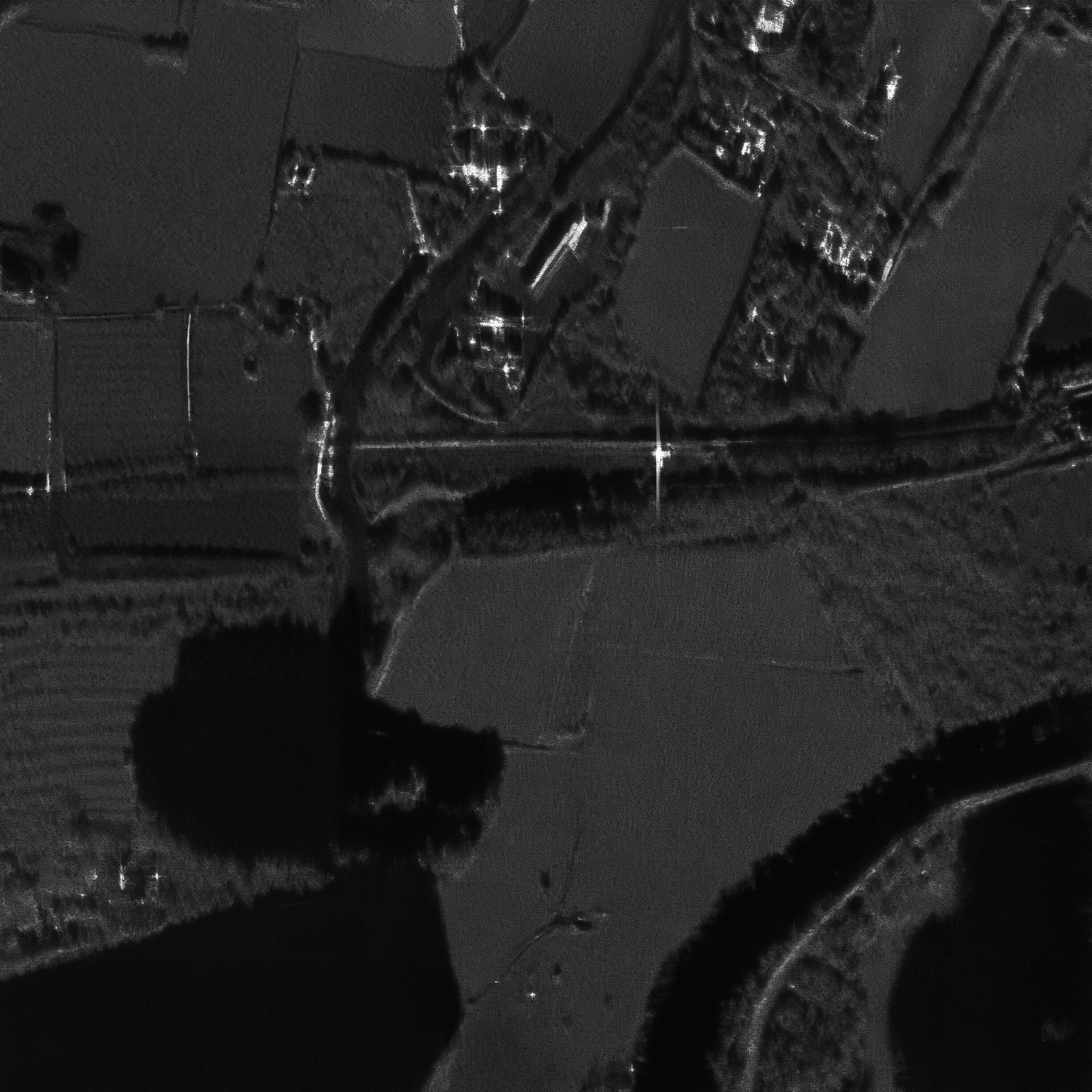}
        \caption{}
    \end{subfigure}
    
    \caption{Enhanced TerraSAR-X satellite images  to aerial resolution  Top row: TerraSAR-X images (512x512, 80cm). Bottom row: AI-enhanced TerraSAR-X images (2048x2048, 40cm).}
    \label{fig:terrasarx}

\end{figure*}

\begin{figure*}
    \centering
    \begin{subfigure}{0.19\textwidth}
        \includegraphics[width=\textwidth]{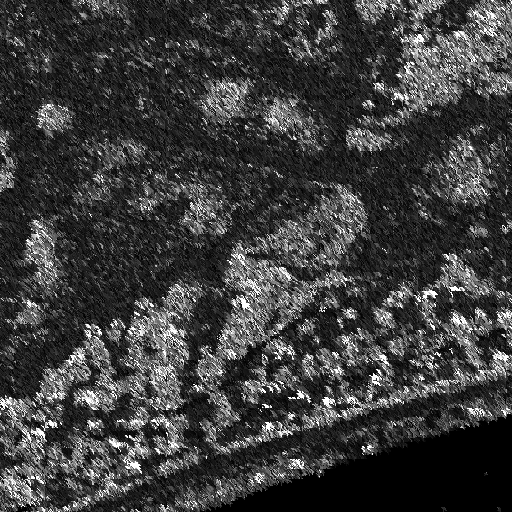}
        \caption{}
    \end{subfigure}\hfill
    \begin{subfigure}{0.19\textwidth}
        \includegraphics[width=\textwidth]{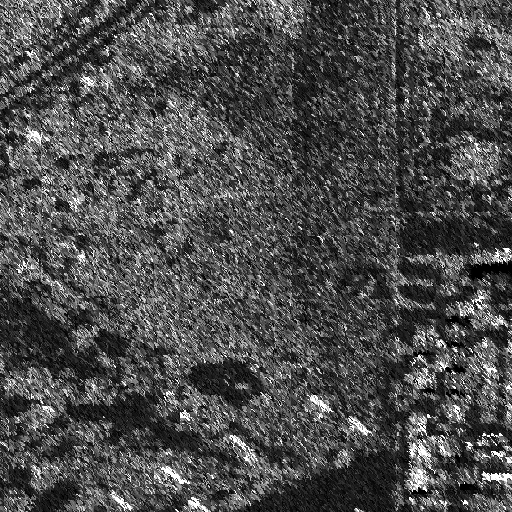}
        \caption{}
    \end{subfigure}\hfill
    \begin{subfigure}{0.19\textwidth}
        \includegraphics[width=\textwidth]{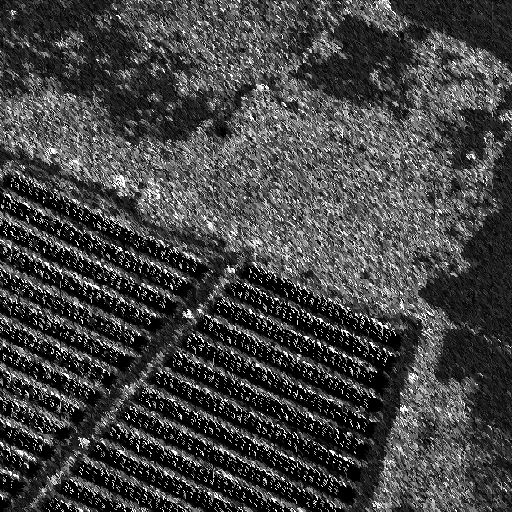}
        \caption{}
    \end{subfigure}\hfill
    \begin{subfigure}{0.19\textwidth}
        \includegraphics[width=\textwidth]{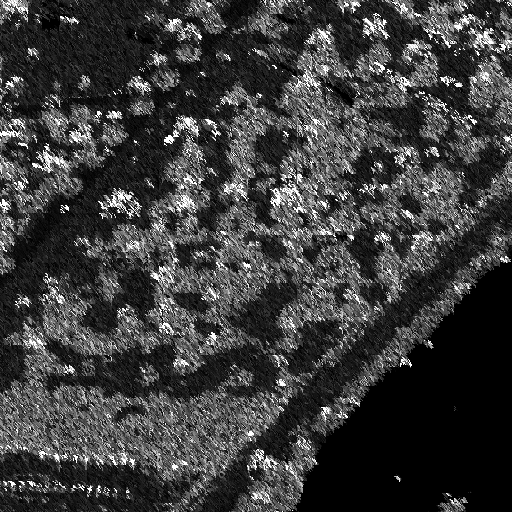}
        \caption{}
    \end{subfigure}\hfill
    \begin{subfigure}{0.19\textwidth}
        \includegraphics[width=\textwidth]{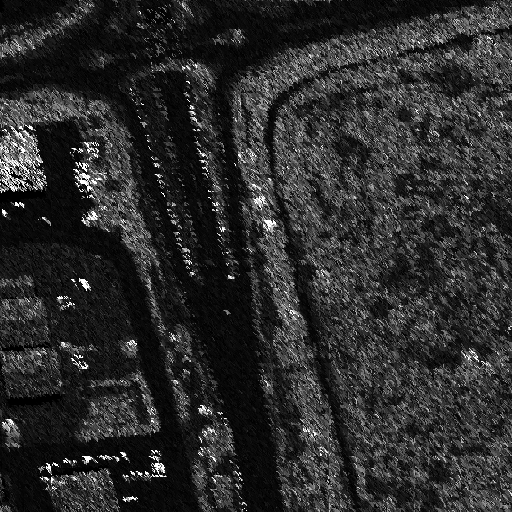}
        \caption{}
    \end{subfigure}

    \vspace{5pt}
    \begin{subfigure}{0.19\textwidth}
        \includegraphics[width=\textwidth]{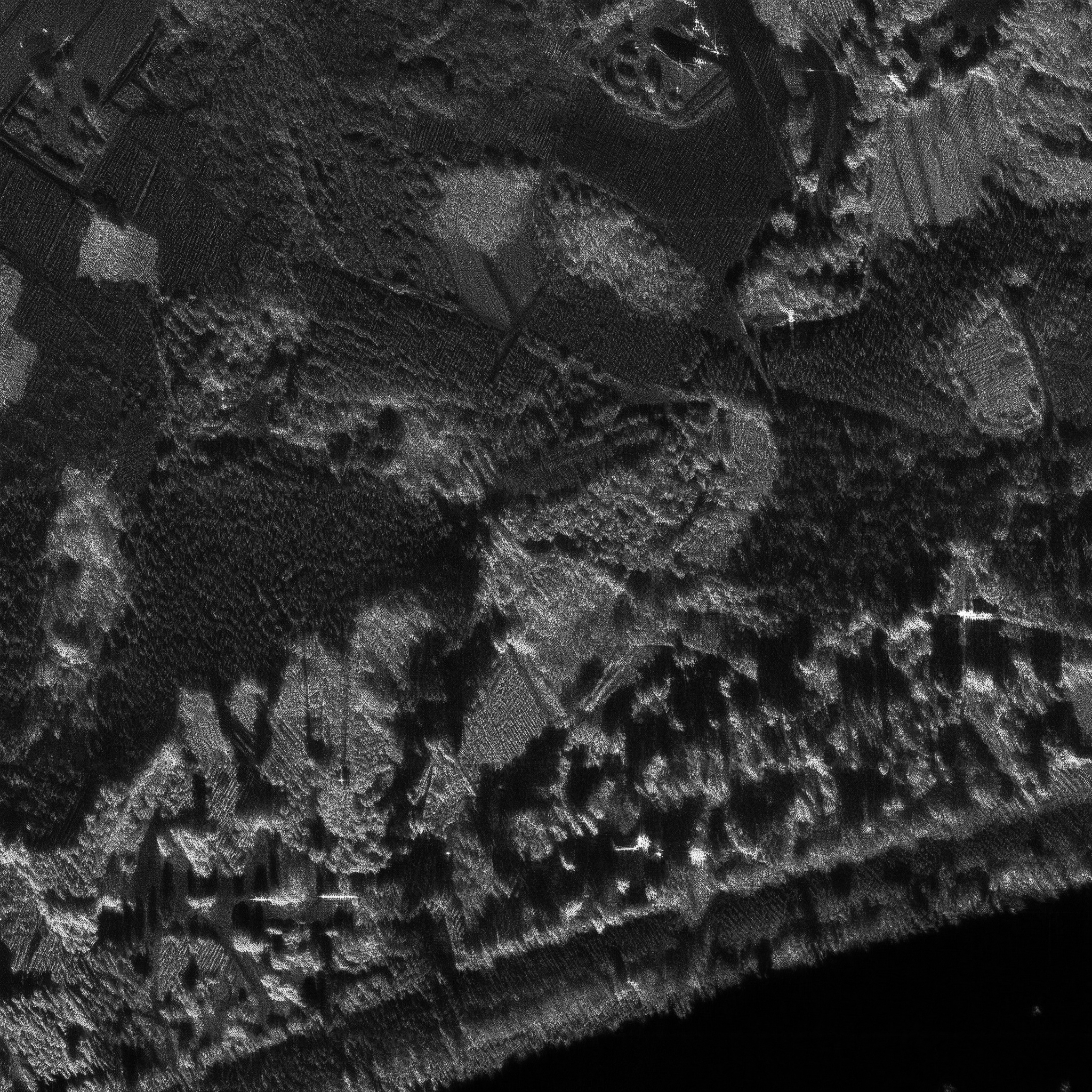}
        \caption{}
    \end{subfigure}\hfill
    \begin{subfigure}{0.19\textwidth}
        \includegraphics[width=\textwidth]{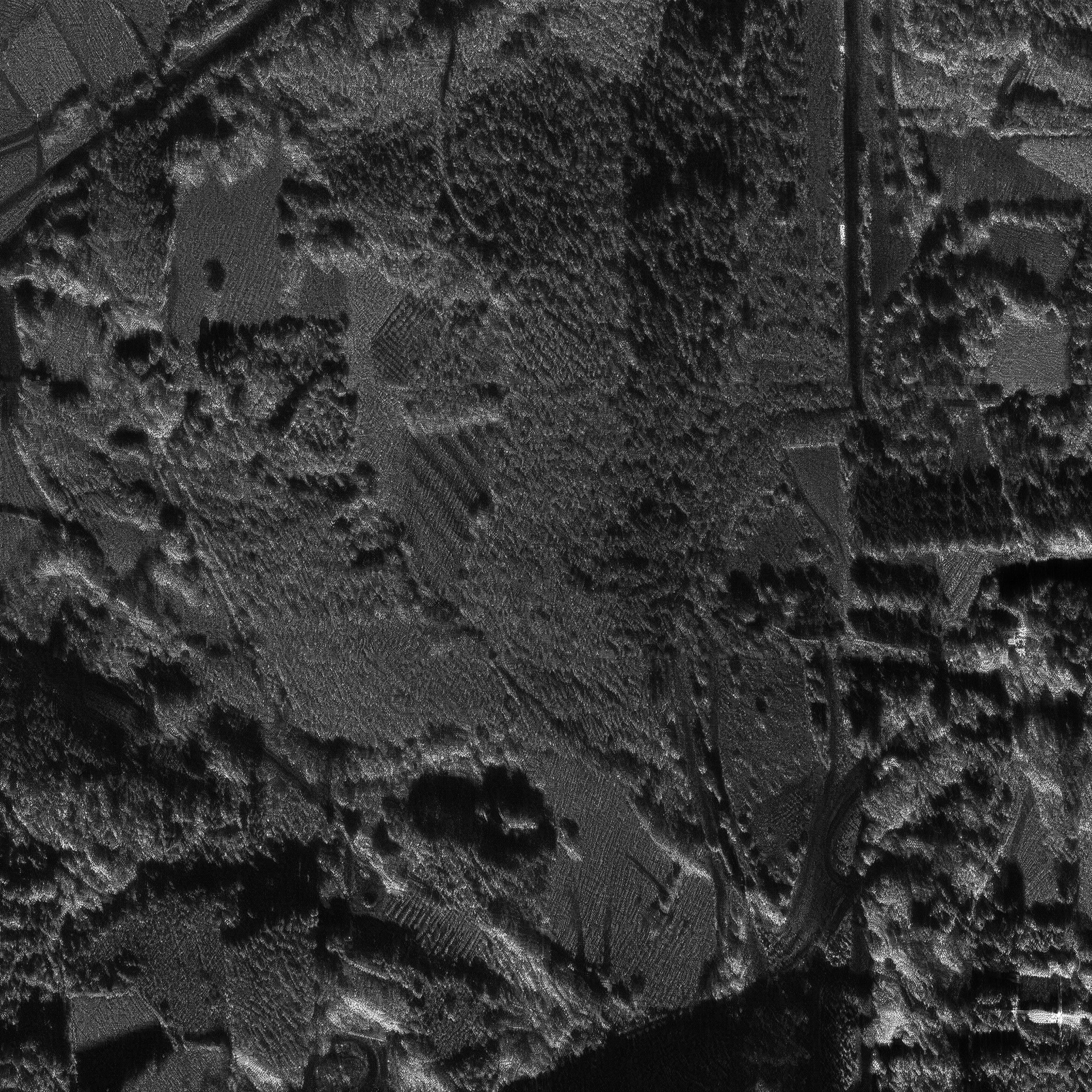}
        \caption{}
    \end{subfigure}\hfill
    \begin{subfigure}{0.19\textwidth}
        \includegraphics[width=\textwidth]{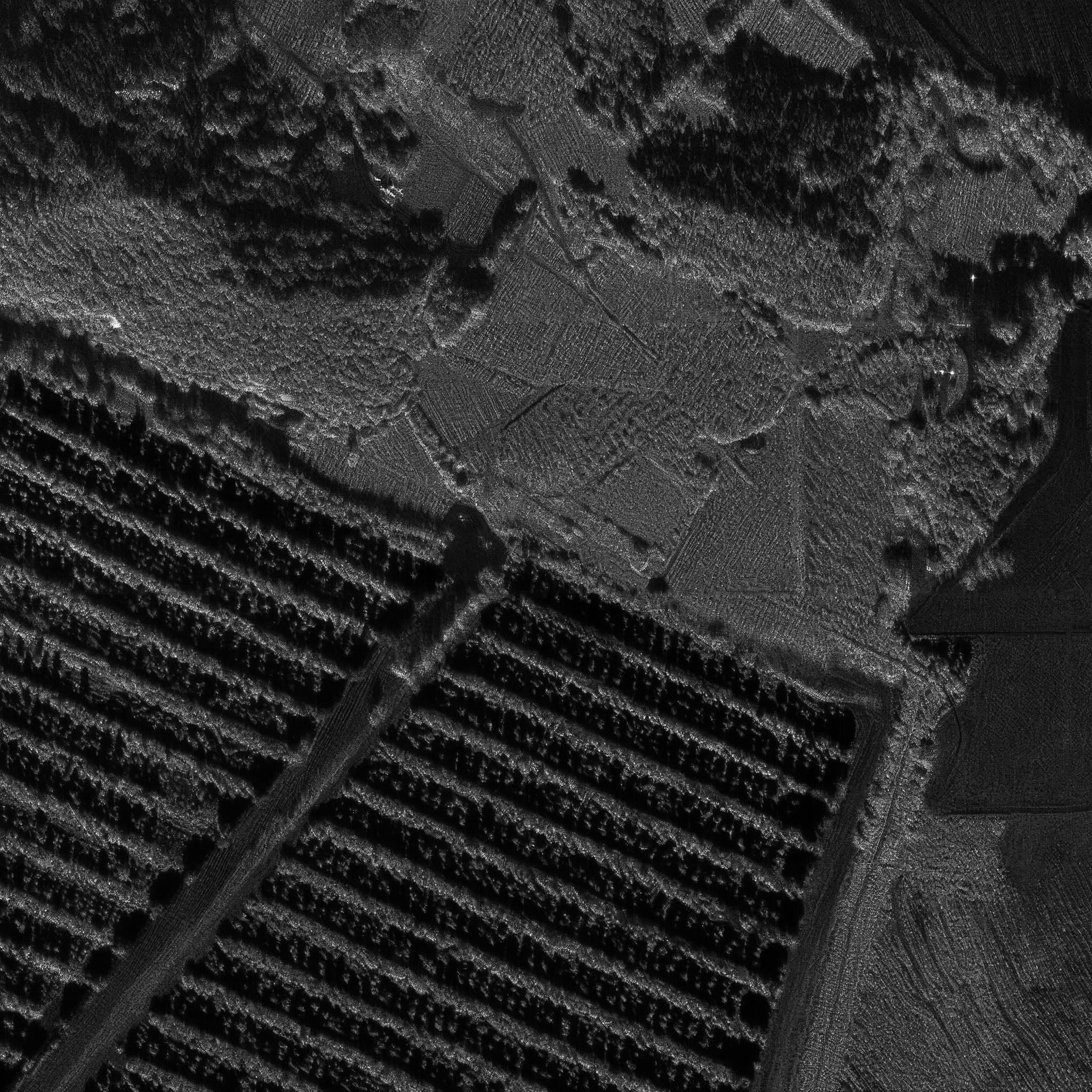}
        \caption{}
    \end{subfigure}\hfill
    \begin{subfigure}{0.19\textwidth}
        \includegraphics[width=\textwidth]{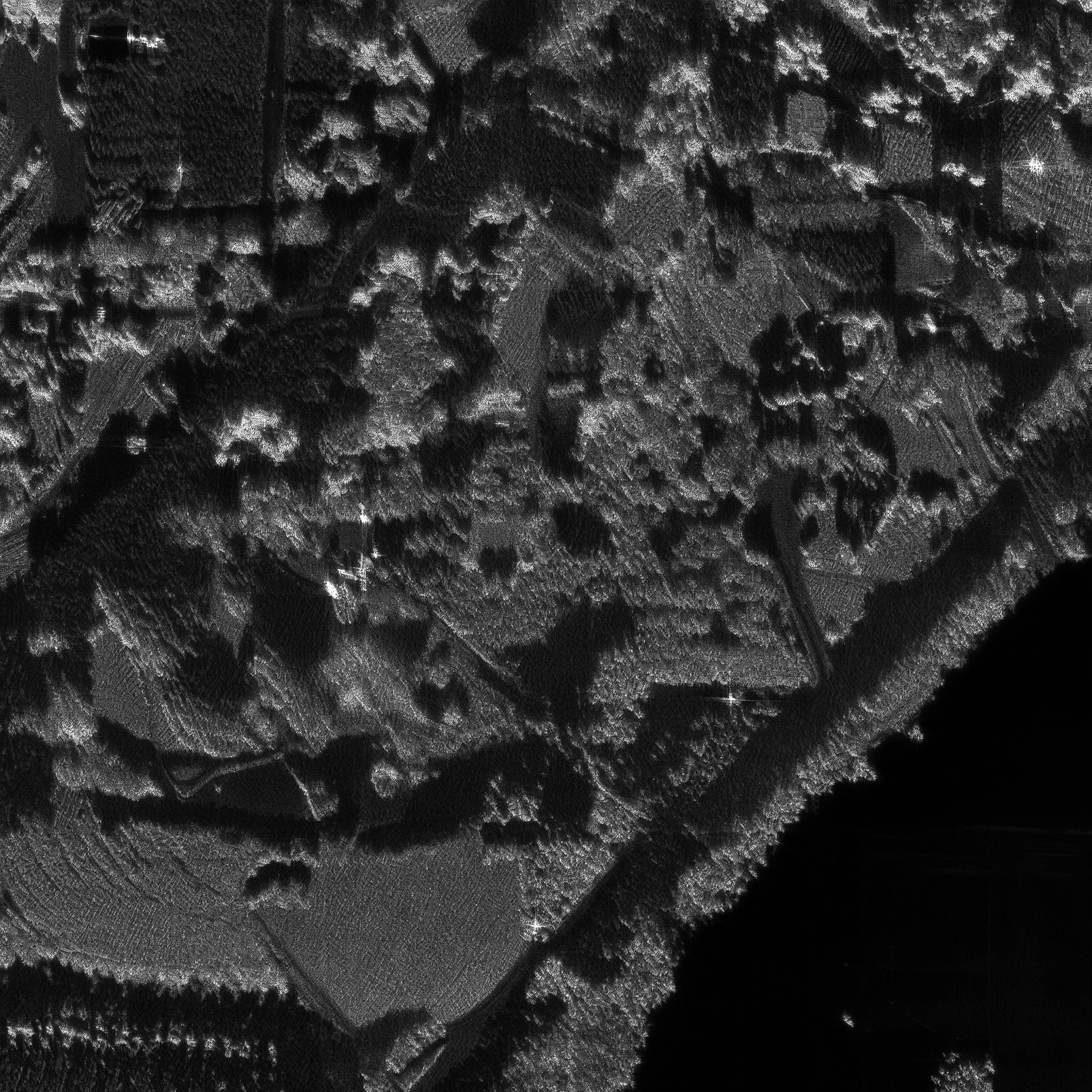}
        \caption{}
    \end{subfigure}\hfill
    \begin{subfigure}{0.19\textwidth}
        \includegraphics[width=\textwidth]{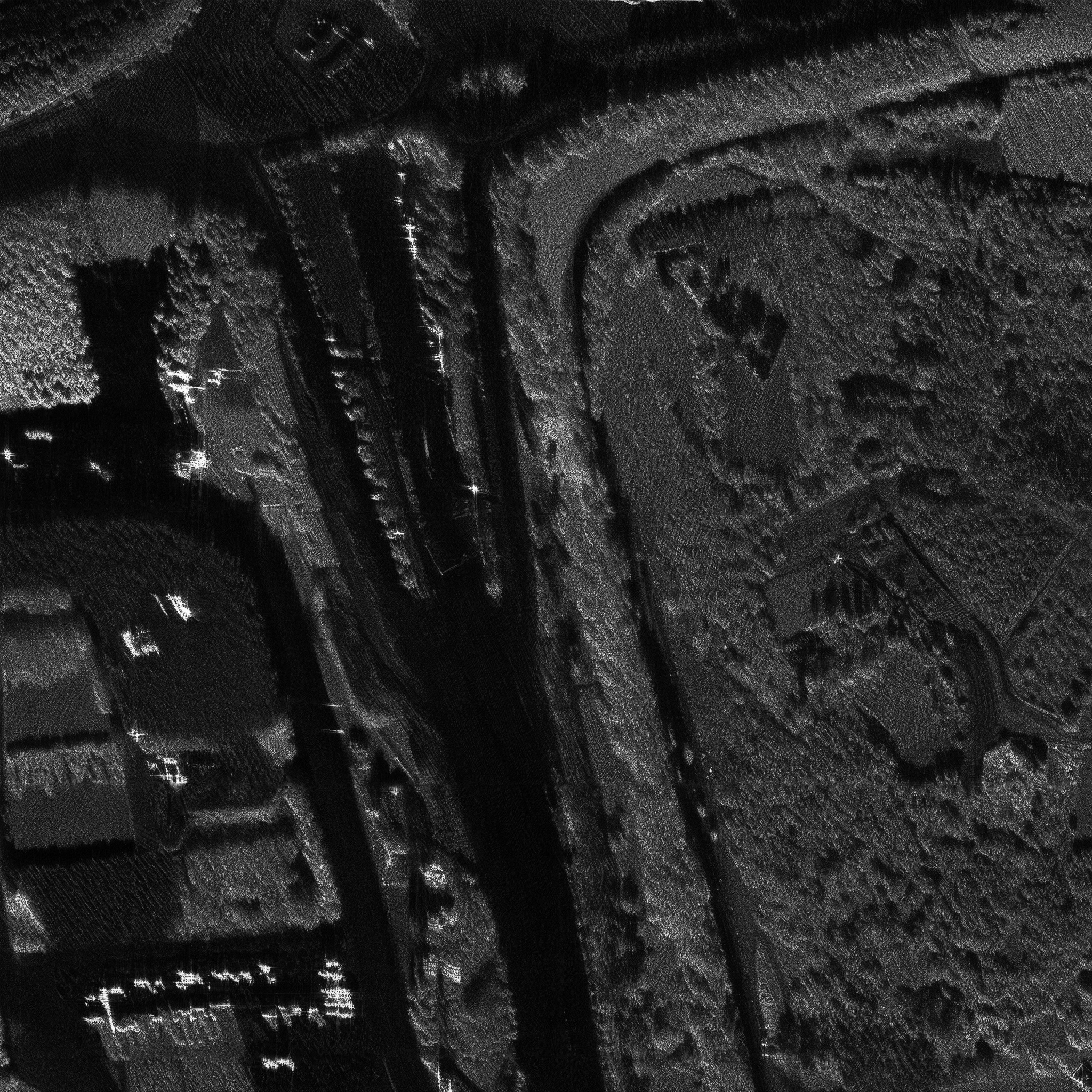}
        \caption{}
    \end{subfigure}
    \caption{Enhanced simulated images from ONERA's Radar simulator with creative content Top row: simulated images (512x512, 40cm). Bottom row: AI-enhanced simulated images (2048x2048, 40cm).}
   \label{fig:emprise}
\end{figure*}

Our pipeline \ref{fig:pipeline} consists in a multi-stage process designed to enhance the resolution of SAR images, such as TerraSAR-X data, by combining latent upscaling with texture refinement through ControlNet modules integrated into Stable Diffusion XL (SDXL) using ComfyUI. We used this pipeline for both TerraSAR-X satellite images \ref{fig:terrasarx} (at 80cm resolution) and simulated images from ONERA's radar  simulator \ref{fig:emprise} (at 40cm resolution). The process begins with a 512×512 input image, which is first upscaled to 1024×1024 using a latent upscaler. The upscaled image is then encoded into the latent space via a VAE encoder and refined by ControlNet (Module 1), which incorporates SDXL Canny and SDXL Tile filters. These filters are configured with end percent values of 0.7 and 0.8, respectively, and a strength of 0.8 to guide the SD80 model (Stable Diffusion XL fine-tuned with a LoRA at a 80cm resolution) in generating a higher-resolution image with an effective ground resolution of 80 cm. This intermediate output is subsequently upscaled to 2048×2048 and processed by ControlNet (Module 2), which also uses the SDXL Canny and SDXL Tile filters, both set with an end percent of 0.8 and a strength of 0.8. This step refines the image through the SD40 model (Stable Diffusion XL fine-tuned with a LoRA at a 40cm resolution) to achieve a final ground resolution of 40 cm. The outputs at each stage are decoded back into the image space using a VAE decoder. This progressive upscaling and refinement approach effectively enhances the realism and resolution of SAR images.

\section{Discussion}
The results presented in both \ref{fig:emprise} and \ref{fig:terrasarx} highlight the promising performance of our pipeline. In our applications, we observe that the speckle noise present in TerraSAR-X satellite images is effectively smoothed by the diffusion process, which acts as a denoiser. Additionally, the generative model successfully introduces realistic details into simulated images that originally lacked fine textures. However, to ensure that the global structure of the image is preserved, the role of ControlNet is crucial. Depending on the strength and the efficiency of the Controlnet, we can see significant modifications to the original structure of the image. This is not in itself a problem if we're indeed looking to generate different SAR images from the same image. A significant challenge in this study concerns the fact that the ControlNet modules used in the upscaling pipeline were originally fine-tuned for controlling optical images, particularly in detecting edges and refining textures. However, SAR images have fundamentally different characteristics, including speckle noise and complex backscatter behavior. This mismatch complicates the ability of pre-trained ControlNet models (optimized for optical data) to effectively condition SAR images. Consequently, this limitation likely impacted the overall performance of the generative pipeline. A key future direction involves fine-tuning ControlNet modules specifically on SAR datasets to better capture the distinct texture and edge patterns inherent in SAR imagery. 

\section{Conclusion}
In conclusion, while our multistage pipeline leveraging ControlNet modules and Stable Diffusion XL has demonstrated promising results to generate new data from scene description and from simulated or TerraSAR-X images. But several challenges remain in fine-tuning ControlNet for SAR data and exploring multimodal conditioning (3D SAR images, sentinel-2, etc.). These efforts will contribute to unlocking new applications for remote sensing and data augmentation.

\bibliographystyle{unsrt} 
\bibliography{references}  

\begin{thebibliography}{1}

\bibitem{Baqu2019SethiR}
R{\'e}mi Baqu{\'e}, Philippe Dreuillet, and H{\'e}l{\`e}ne~M. Oriot.
\newblock Sethi : Review of 10 years of development and experimentation of the remote sensing platform.
\newblock {\em 2019 International Radar Conference (RADAR)}, 2019.

\bibitem{empriseem_ai_images}
DEMR-SEM ONERA.
\newblock Examples of ai generated sar images, 2024.

\bibitem{ho2020denoising}
Jonathan Ho, Ajay Jain, and Pieter Abbeel.
\newblock Denoising diffusion probabilistic models, 2020.

\bibitem{goodfellow2014generative}
Ian~J. Goodfellow, Jean Pouget-Abadie, Mehdi Mirza, Bing Xu, David Warde-Farley, Sherjil Ozair, Aaron Courville, and Yoshua Bengio.
\newblock Generative adversarial networks, 2014.

\bibitem{rombach2022highresolution}
Robin Rombach, Andreas Blattmann, Dominik Lorenz, Patrick Esser, and Björn Ommer.
\newblock High-resolution image synthesis with latent diffusion models, 2022.

\bibitem{trouve2024synthesis}
Nicolas Trouve, Nathan Letheule, Olivier Leveque, Ilias Rami, and Elise Colin.
\newblock Sar image synthesis using text conditioned pre-trained generative ai models.
\newblock In {\em Proceedings of EUSAR 2024; 15th European Conference on Synthetic Aperture Radar}, Munich, Germany, 2024. VDE, VDE,ITG.

\bibitem{debuysere}
Debuysere Solene, Trouve Nicolas, Letheule Nathan, Leveque Olivier, and Colin Elise.
\newblock Synthesizing sar images with generative ai: Expanding to large-scale imagery.
\newblock In {\em Proceedings of RADAR 2024}, Rennes, France, 2024.

\bibitem{podell2023sdxl}
Dustin Podell, Zion English, Kyle Lacey, Andreas Blattmann, Tim Dockhorn, Jonas Müller, Joe Penna, and Robin Rombach.
\newblock Sdxl: Improving latent diffusion models for high-resolution image synthesis, 2023.

\bibitem{zhang2023adding}
Lvmin Zhang, Anyi Rao, and Maneesh Agrawala.
\newblock Adding conditional control to text-to-image diffusion models, 2023.

\end{thebibliography}

\end{document}